\documentclass[journal=jctcce,manuscript=article]{achemso}

\usepackage[utf8]{inputenc}
\usepackage[english]{babel}
\usepackage{graphicx}
\usepackage{amsmath, amssymb}
\usepackage[dvipsnames]{xcolor}
\usepackage{booktabs, multirow}
\usepackage{soul}
\usepackage{url}
\usepackage[version=4]{mhchem}

\author{Arthur Hagopian}
\affiliation[ICGM]{ICGM, Univ Montpellier, CNRS, ENSCM, Montpellier, France}
\alsoaffiliation[RS2E]{R\'eseau sur le Stockage Electrochimique de l’Energie (RS2E), FR CNRS 3459, Hub de l'Energie, Amiens, France}
\author{Marie-Liesse Doublet}
\affiliation[ICGM]{ICGM, Univ Montpellier, CNRS, ENSCM, Montpellier, France}
\alsoaffiliation[RS2E]{R\'eseau sur le Stockage Electrochimique de l’Energie (RS2E), FR CNRS 3459, Hub de l'Energie, Amiens, France}
\author{Jean-S\'ebastien Filhol}
\affiliation[ICGM]{ICGM, Univ Montpellier, CNRS, ENSCM, Montpellier, France}
\alsoaffiliation[RS2E]{R\'eseau sur le Stockage Electrochimique de l’Energie (RS2E), FR CNRS 3459, Hub de l'Energie, Amiens, France}
\author{Tobias Binninger}
\email{tobias.binninger.science@gmx.de}
\affiliation[ICGM]{ICGM, Univ Montpellier, CNRS, ENSCM, Montpellier, France}
\alsoaffiliation[RS2E]{R\'eseau sur le Stockage Electrochimique de l’Energie (RS2E), FR CNRS 3459, Hub de l'Energie, Amiens, France}

\title{Advancement of the Homogeneous Background Method for the Computational Simulation of Electrochemical Interfaces}

\begin{document}



\begin{abstract}
Computational studies of electrochemical interfaces based on density-functional theory (DFT) play an increasingly important role in present research on electrochemical processes for energy conversion and storage. The homogeneous background method (HBM) offers a straightforward approach to charge the electrochemical system within DFT simulations, but it typically requires the specification of the ``active'' fraction of excess electrons based on a certain choice of the electrode--electrolyte boundary location, which can be difficult in presence of electrode-surface adsorbates or explicit solvent molecules. In this work, we present a methodological advancement of the HBM, both facilitating and extending its applicability. The advanced version neither requires energy corrections nor the specification of the ``active'' fraction of excess electrons, providing a versatile and readily available method for the simulation of charged interfaces also when adsorbates or explicit solvent molecules are present. Our computational DFT results for Pt(111), Au(111) and Li(100) metal electrodes in high-dielectric-constant solvents demonstrate an excellent agreement in the interfacial charging characteristics obtained from simulations with the advanced HBM in comparison with the (linearized) Poisson-Boltzmann model (PBM). 
\end{abstract}

\section{Introduction}

The decarbonization of energy economy and industry will require efficient electrochemical processes for energy conversion and storage. Battery technologies, e.g., are already massively applied in the e-mobility sector~\cite{2019_IEEE_Adib} and ``green'' hydrogen, produced \textit{via} water electrolysis, will provide chemical feedstock, e.g., for the direct-reduced-iron (DRI) process~\cite{2021_Joule_Fan} in the steel industry. Therefore, enormous research efforts are currently devoted to the investigation and improvement of the microscopic processes occurring at charged electrochemical interfaces. Enabled by the advancements in high-performance computing (HPC) technology, atomistic computational simulations have evolved into an indispensable part of the research in this field~\cite{2009_NatChem_Greeley, 2017_JAmChemSoc_Goddard, 2021_NatCatal_Koper}. 

Density-functional theory (DFT) provides an efficient and well-established framework for the \textit{ab initio} simulation of the electrode side of an electrochemical interface. However, the adequate treatment of the electrolyte side, interfacial charging, and electrode potential variations imposes a major challenge for DFT-based simulations. All-explicit simulations of both electrode and electrolyte~\cite{2018_JChemPhys_Gross, 2019_JPhysChemLett_Sprik, 2021_JACSAu_Le_Iannuzzi} are generally still prohibitively expensive in terms of computational resources. Therefore, different methods have been developed for an implicit treatment of both the solvent and the ionic charge of the electrolyte. The implicit solvent is typically described at the level of polarizable continuum models (PCM)~\cite{2012_JChemPhys_Marzari, VASPsol2014-Dielectric, 2014_PhysRevB_Andreussi_Marzari, 2018_JChemPhys_Schwarz}, but also advanced methods are used such as the reference interaction site model (RISM)~\cite{1999_JChemPhys_Kovalenko_Hirata, 2017_PhysRevB_Otani, 2019_ACSApplMaterInterf_Eikerling, 2021_JPhysCondMat_Tesch_Kowalski_Eikerling}.  For the ionic charge of the electrolyte, various implicit models have been developed, including the homogeneous compensating background~\cite{2006_AngewandteChem_Filhol}, Gaussian distributions~\cite{2019_JChemPhys_Hoermann_Marzari}, and screening layers described by Poisson-Boltzmann-type equations~\cite{2006_PhysRevB_Otani, 2008_PhysRevB_Jinnouchi, 2010_arXiv_Dabo, 2012_PhysRevB_Letchworth-Weaver_Arias, 2017_JChemPhys_Sundararaman_Arias, 2019_JChemPhys_Mathew_Hennig, 2019_JChemPhys_Nattino_Marzari,
2019_JChemPhys_Melander_Honkala}. The latter approaches are formalized by a general joint density-functional theory (JDFT)~\cite{2005_JPhysChemB_Petrosyan_Arias, 2012_PhysRevB_Letchworth-Weaver_Arias} that provides a combined DFT description of the electron density together with the ionic and dielectric densities of the electrolyte. It has been noted, however, that the level of complexity of the ionic counter-charge model had rather little influence on the numerically obtained value of the interface capacitance, which has been attributed to the dominating role of the underlying dielectric solvent model and the corresponding definition of the solvent boundary~\cite{2019_JChemPhys_Melander_Honkala}. 
 
To model a charged interface within DFT-based simulations, the electron number of the simulation cell, which contains the electrode slab embedded in solvent, is changed with respect to the electron number of the neutral electrode slab. In order to preserve overall charge neutrality, a counter-charge distribution must be included that compensates the electronic excess charge and mimics the ionic charge of the electrolyte in the electrochemical double-layer. The homogeneous background method (HBM)~\cite{2006_AngewandteChem_Filhol, 2011_PhysChemChemPhys_Filhol, 2020_PCCP_Kopac} utilizes a compensating homogeneous background charge, $\rho_{bg} = \mathrm{const}$, across the entire simulation cell, which generally is a ``built-in'' functionality of periodic DFT-codes. The uniform counter-charge distribution must be regarded as rather unphysical from the perspective of an electrochemical interface, in particular because it extends across the bulk of the electrode slab. This aspect complicates the application of the HBM for two reasons: First, an \textit{a posteriori} energy correction is typically required, and second, the fraction of ``active'' excess electrons at the electrode surface must be estimated based on a certain choice of the electrode--electrolyte boundary location~\cite{2006_AngewandteChem_Filhol, 2011_PhysChemChemPhys_Filhol, 2020_PCCP_Kopac}. The latter aspect, in particular, makes the use of the HBM difficult in cases where the definition of the electrode--electrolyte boundary becomes ambiguous, e.g. due to the presence of surface adsorbate species or explicit solvent molecules that become, to a certain extent, ``metallized''~\cite{2020_PCCP_Kopac} and therefore cannot be uniquely ascribed to either the electrode or the electrolyte side. In the present work, we resolve this shortcoming by developing an advanced version of the HBM that can be straightforwardly applied without the typically employed correction schemes and without requiring the knowledge of the ``active'' fraction of excess electrons, thus significantly extending the applicability of the HBM.

As a computational validation, we compare the advanced HBM with the linearized Poisson-Boltzmann model (PBM)~\cite{2019_JChemPhys_Mathew_Hennig}, both in combination with a PCM for the implicit solvent as implemented in the VASPsol package~\cite{VASPsol2014-Dielectric}. The PBM employs a physically inspired counter-charge distribution based on the Gouy-Chapman theory for the electrochemical double-layer~\cite{2010_book_Schmickler}, where the ion densities $n_i(\mathbf{r}) = n_{i,0}\exp\left(-z_i e\phi(\mathbf{r})/k_{\mathrm{B}}T\right)$ in the electrolyte are determined by the local electrostatic potential $\phi(\mathbf{r})$ via Boltzmann statistics. In the linearized version, an ionic charge density $\rho_{\text{ion}}(\mathbf{r}) = -\epsilon_0 \epsilon_{\text{r}}\kappa^2(\phi(\mathbf{r})-\phi_{\mathrm{elyte}})$ is obtained, where $\epsilon_{\text{r}}$ is the relative permittivity of the electrolyte, $\kappa = 1/\lambda_{\text{D}}$ is the inverse Debye length, and $\phi_{\mathrm{elyte}}$ is the reference electrostatic potential in the bulk electrolyte. Combined with the dielectric charge density of the linear PCM solvent, this leads to the linearized Poisson-Boltzmann equation $\nabla[\epsilon_{\text{r}}\nabla\phi]-\epsilon_{\text{r}}\kappa^2(\phi-\phi_{\mathrm{elyte}}) = -(\rho_{\text{ext}}+\rho_{\text{e}})/\epsilon_0$, the solution of which produces an ionic counter-charge distribution $\rho_{\text{ion}}(\mathbf{r})$ with an essentially exponential decay, as $\exp(-z/\lambda_{\text{D}})$, from the electrolyte boundary towards the bulk of the electrolyte. However, it should be noted that, strictly speaking, the mean-field thermodynamics underlying the PBM do not apply to the description of the atomistic interactions at an electrochemical interface~\cite{gauthier_challenges_2019}. In spite of their very different counter-charge distributions, we demonstrate that both the HBM and PBM yield essentially equal results for the free energy and capacitance in DFT-based simulations of charged electrochemical interfaces. We rationalize this finding by an electrostatic model, highlighting the central role of the gap region~\cite{2012_PhysRevB_Letchworth-Weaver_Arias, 2015_JChemTheoComp_Filhol, 2019_JChemPhys_Hoermann_Marzari} between the electrode surface and the boundary of the implicit solvent region in determining the interface capacitance for both models.

\section{Theory and Methods}

\subsection{Grand potential, electrode potential, and capacitance}

We begin with a brief review of the general thermodynamic equations for an electrochemical interface~\cite{2010_book_Schmickler, 2019_JChemPhys_Melander_Honkala}, which hold, in particular, for the PBM method. Their validity for the HBM, however, must be carefully assessed as discussed in the subsequent section. Electrochemical processes involve the exchange of electrons and ions between the interface region and the bulk of the electrode and electrolyte, respectively. The interface therefore corresponds to an open system coupled to an electron reservoir, the electrode, and an ion reservoir, the electrolyte. The corresponding grand canonical ensemble is characterized by a certain electron chemical potential $\mu_e$ and ion chemical potentials $\mu_i$ for each of the ion species present in the electrolyte.\footnote{In a complete description~\cite{2010_book_Schmickler}, also the solvent molecules must be included. However, they do not influence the present discussion, so we neglect this aspect for the sake of clarity.} The equilibrium state of the electrochemical interface corresponds to the minimum of the grand potential
\begin{align} 
\label{eq_grand_pot_general}
\Omega \,=\, A - \mu_e N_e - \sum_i \mu_i N_i
\end{align}
where $A$ is the Helmholtz free energy, and $N_e$ and $N_i$ are the \emph{excess} particle numbers of electrons and ions, respectively, which quantify the difference with respect to the uncharged interface.

An aspect of central importance is the overall charge neutrality. The electronic excess charge at the electrode surface must, on average, be balanced by the ionic excess charge of the electrolyte to avoid energy divergence. Therefore, the electron and ion numbers $N_e$ and $N_i$ cannot vary independently, but they must fulfill the charge neutrality condition $N_e = \sum_i z_i N_i$ with the charge numbers $z_i$ of the ion species. Splitting the ion chemical potentials as $\mu_i = \mu_i^0 + e z_i \phi_{\mathrm{elyte}}$, where $\phi_{\mathrm{elyte}}$ is the inner electrostatic potential in the bulk electrolyte, we obtain the grand potential in the form
\begin{align}
\Omega\, & =\, A - (\mu_e + e\,\phi_{\mathrm{elyte}}) N_e - \sum_i \mu_i^0 N_i \label{eq_grand_pot_PBM_combined}
\end{align}
where we used the charge neutrality condition. 

The electrode potential $\Phi$ can be defined as a ``work function'' in the electrolyte environment~\cite{2012_PhysRevB_Letchworth-Weaver_Arias},
\begin{align}
e\,\Phi \,=\, (-e)\phi_{\mathrm{elyte}} - \mu_e
\label{eq_electrode_potential}
\end{align}
where the electron chemical potential $\mu_e$ is referenced to the electrostatic potential energy $(-e)\,\phi_{\mathrm{elyte}}$ in the bulk electrolyte. The (differential) interface capacitance
\begin{align}
\frac{1}{C} \,=\, \left(\frac{\partial \Phi}{\partial q}\right)_{T,V}
\label{eq_capacitance_Phi_q_PBM}
\end{align}
relates changes in the interfacial charge $q$ to changes in the electrode potential $\Phi$. Since we consider fixed temperature $T$ and volume $V$ throughout the following, we omit the indication of the $T,V$-subscript. Due to the charge neutrality requirement discussed above, the electron and ion numbers cannot be varied independently. We therefore consider variations in electron number $N_e$, and thus interfacial charge $q$, under fixed ion chemical potentials $\mu_i$. In this \emph{partial} grand canonical setting~\cite{2021_PhysRevB_Binninger}, a variation in $N_e$ implicitly includes a compensating variation in the ion numbers $N_i$ to maintain charge neutrality. The fixed ion chemical potentials $\mu_i = \mu_i^0 + e z_i \phi_{\mathrm{elyte}}$ also fix the electrostatic potential $\phi_{\mathrm{elyte}}$ in the electrolyte. Therefore, with the electrode potential given in Eq.~\eqref{eq_electrode_potential}, the interface capacitance of Eq.~\eqref{eq_capacitance_Phi_q_PBM} fulfills
\begin{align}
\frac{e^2}{C} \,=\, \left(\frac{\partial \mu_e}{\partial N_e}\right)_{\{\mu_i\}}
\label{eq_capacitance_F_N_PBM}
\end{align}
where we used the relation $q = -e N_e$ for the interfacial charge, which holds in the PBM, but not in the HBM, as discussed below. From now on, we consider all partial derivatives under constant ion chemical potentials $\{\mu_i\}$, and, for simplicity, we omit the indication of the $\{\mu_i\}$-subscript. From the well-known relation
\begin{align}
\frac{\partial\Omega}{\partial \mu_e} \,=\,  - N_e
\label{eq_deriv1_Omega_mu_PBM}
\end{align}
we find
\begin{align}
\frac{\partial^2\Omega}{\partial \mu_e^2} \,=\, -\frac{\partial N_e}{\partial \mu_e} \,=\, -\frac{C}{e^2}
\label{eq_deriv2_Omega_mu_PBM}
\end{align}
where we used Eq.~\eqref{eq_capacitance_F_N_PBM} and $\left(\partial N_e/\partial \mu_e\right) = \left(\partial \mu_e/\partial N_e\right)^{-1}$. Due to the fixed electrolyte potential $\phi_{\mathrm{elyte}}$, we have $\partial/\partial \Phi = (-e)\partial/\partial \mu_e$ for the derivative with respect to the electrode potential $\Phi$ according to Eq.~\eqref{eq_electrode_potential}. Consequently,
\begin{align}
\frac{\partial^2 \Omega}{\partial \Phi^2} \,=\, -C
\label{eq_capacitance_Omega_Phi_PBM}
\end{align}
which is the well-known relation that the curvature of the grand potential as a function of the electrode potential is equal to the negative of the capacitance.

\subsection{Grand potential and capacitance in HBM}

The original HBM~\cite{2006_AngewandteChem_Filhol} adopts an electronic perspective on the definition of electrode potential and grand potential. Accordingly, a number of corrections are typically applied to correct for the unphysical influence of the homogeneous background charge~\cite{2006_AngewandteChem_Filhol, 2011_PhysChemChemPhys_Filhol, 2020_PCCP_Kopac}. Motivated by the general form of the grand potential~\eqref{eq_grand_pot_general}, we explore the consequences of a novel perspective, including the homogeneous background charge in the definition of the grand potential within the HBM. In this way, we regard the homogeneous background charge not only as a numerical tool for avoiding the energy divergence of infinite charged systems, but we consider it as the simplest possible model for a physical compensating counter-charge distribution. Since the homogeneous background extends across both the electrolyte region and the electrode slab, the HBM essentially combines two different models: First, the homogeneous background charge in the electrolyte region models the ionic counter charge of the electrochemical double-layer. And second, the homogeneous background charge in the electrode slab region can be considered as an entirely delocalized model for a dopant of the electrode material. Therefore, charging the electrode--electrolyte system with the HBM consists of both charging the electrochemical double-layer and changing the dopant concentration of the electrode at the same time. We will see below that, in fact, the latter dopant contribution to the grand potential is negligible for the case of metal electrodes, making the HBM an easy-to-use method for such systems.

The chemical potential of the HBM counter-charge (i.e the uniform background) is given by its electrostatic interaction with the rest of the system as $\mu_{bg} = e\bar{\phi}$ with the mean electrostatic potential $\bar{\phi}$ of the simulation cell~\cite{2011_PhysChemChemPhys_Filhol, 2020_PCCP_Kopac}, and the corresponding ``background particle number'' $N_{bg} = q_{bg}/e$ is simply derived from the background charge $q_{bg}$. Including this contribution in the Legendre transformation, analogous to Eq.~\eqref{eq_grand_pot_general}, the HBM grand potential reads
\begin{align} 
\label{eq_grand_pot_HBM}
\Omega \, & =\, A - \mu_e N_e - \mu_{bg} N_{bg} \nonumber \\
& =\, A - (\mu_e + e\bar{\phi})N_e
\end{align}
where we used the charge neutrality condition $N_{bg} = N_e$ (note that $N_e$ only refers to the \emph{excess} electron number with respect to the uncharged electrode). Because the global system is neutral, this expression does \emph{not} depend on the choice of a certain potential reference.

The definition of the electrode potential $\Phi$ according to Eq.~\eqref{eq_electrode_potential} remains valid also within the HBM. The definition of the interface capacitance within HBM, however, is less straightforward. In principle, we still assume that we can define a ``true'' interface capacitance according to Eq.~\eqref{eq_capacitance_Phi_q_PBM}. However, the interfacial charge $q$ appearing in this expression is \emph{not} anymore equal to $-e N_e$ with the total excess electron number $N_e$ of the simulation cell. Because the homogeneous background charge extends across the bulk region of the metal electrode, a certain part $N_b$ of the excess electrons will screen the background charge therein and therefore \emph{not} contribute to the electrode surface charging. Only the remaining part $N_s = N_e - N_b$ of the excess electrons results in a charging of the electrode--electrolyte interface. The interfacial charge $q = -e N_s$ in the definition of the interface capacitance according to Eq.~\eqref{eq_capacitance_Phi_q_PBM} is therefore not clearly defined within the HBM \textit{a priori}. 

Nevertheless, we can still define a capacitance analogous to Eq.~\eqref{eq_capacitance_Phi_q_PBM} simply using the charge $q_e = -e N_e$ of the total number of excess electrons,
\begin{align}
\frac{1}{C_{\Phi N}} \,=\, -\frac{1}{e}\,\frac{\partial \Phi}{\partial N_e}
\label{eq_capacitance_Phi_N_HBM}
\end{align}
where we have to keep in mind that this capacitance will not be equal to the ``true'' interface capacitance, as discussed above. Also, in analogy to Eq.~\eqref{eq_capacitance_Omega_Phi_PBM}, we define the capacitance
\begin{align}
C_{\Omega \Phi} \,=\, -\frac{\partial^2 \Omega}{\partial \Phi^2}
\label{eq_capacitance_Omega_Phi_HBM}
\end{align}
where $\Omega$ refers to the HBM grand potential of Eq.~\eqref{eq_grand_pot_HBM}. 

We saw in the previous section that the capacitance definitions according to Eqs.~\eqref{eq_capacitance_Phi_N_HBM} and \eqref{eq_capacitance_Omega_Phi_HBM} are precisely equivalent in the PBM and simply equal to the interface capacitance of Eq.~\eqref{eq_capacitance_Phi_q_PBM}. In the HBM, however, $C_{\Phi N}$ and $C_{\Omega \Phi}$ correspond to different quantities. As discussed above, $C_{\Phi N}$ is expected to be different from the ``true'' HBM interface capacitance as it is including the contribution from the excess electrons in the electrode bulk region. The central question arises whether at least $C_{\Omega \Phi}$ can provide a representation of the latter, to be addressed in detail in the following.

\section{Computational Details}

Calculations were performed within the density-functional theory (DFT) framework, using the Vienna ab initio simulation package (VASP)~\cite{kresse_ab_1993}. The electronic wave functions were expanded in a plane-wave basis set with a kinetic energy cutoff up to $450\,\mathrm{eV}$ for Li and $550\,\mathrm{eV}$ for Pt and Au systems. Projector augmented wave (PAW) pseudopotentials as implemented in VASP were used~\cite{perdew_generalized_1996-1}. Exchange-correlation effects were accounted for by the generalized gradient approximation (GGA) using the functional of Perdew, Burke and Ernzerhof (PBE)~\cite{perdew_generalized_1996-1}. The convergence criterion was set to $10^{-6}\,\mathrm{eV}$ for the electronic self-consistent iterations. Structural relaxation was performed until the maximum force on any atom was below $10^{-2}\,\mathrm{eV\,\AA^{-1}}$. Periodic interface calculations were performed on slabs comprising 7 atomic layers generated through the cleavage of the relaxed bulk crystals along the (100)-surface orientation for Li and (111)-surface orientation for Pt and Au. Slab surfaces comprised 1 surface atom ($11.84\,\mathrm{\AA^2}$ surface area), 4 surface atoms ($27.28\,\mathrm{\AA^2}$ surface area) and 4 surface atoms ($29.13\,\mathrm{\AA^2}$ surface area) for Li (100), Pt (111) and Au (111), respectively. To prevent inhomogeneous charging of the two surfaces of the slab, cells were built symmetric in the \textit{z}-direction and the central atomic layer was kept frozen while the other layers were allowed to relax. Except when stated otherwise, a $15\,\mathrm{\AA}$-wide interspace region separated the periodic images of the slab in the \textit{z}-direction. Brillouin-zone sampling was done according to Monkhorst-Pack~\cite{monkhorst_special_1976} on a $14 \times 14 \times 1$ $\Gamma$-centered $k$-point grid for Pt and Li, and $11 \times 11 \times 1$ for Au. The surrounding electrolyte within the interspace region was described by an implicit solvent using the polarizable continuum model (PCM) as implemented in VASPsol~\cite{VASPsol2014-Dielectric}. The dielectric constant of ethylene carbonate ($\epsilon_b$ = 89.9), which is a major solvent component of Li-battery electrolytes, was used for Li calculations, while the one of water ($\epsilon_b$ = 78.4) was used for Pt and Au. The effective surface tension parameter $\tau$ was set to $0\,\mathrm{eV\,\AA^{-2}}$~\cite{gauthier_challenges_2019} and the critical density parameter $n_c$ was set to $5\times 10^{-5}\,\mathrm{\AA^{-3}}$ for Li~\cite{hagopian_ab_2021} and $2.5\times 10^{-3}\,\mathrm{\AA^{-3}}$ for Pt and Au~\cite{2019_JChemPhys_Mathew_Hennig}. The cavity-shape parameter $\sigma$ was set to the default value of $0.6$. Unless otherwise stated, for charged interface calculations with the Poisson-Boltzmann model (PBM) from VASPsol, a Debye length $\lambda_{\mathrm{D}}$ of $1.5\,\mathrm{\AA}$ ($1/10$ of the total interspace width) was used in order to ensure sufficient accommodation of the counter-charge distribution within the interspace region. To compute the bulk chemical hardness $\eta_b^0$ of Pt, calculations were performed on a Pt-bulk cell that was constructed by removing the interspace region and one surface atomic layer from the Pt(111)-slab cell used for interface calculations, and the $k$-point grid was increased to $20 \times 20 \times 10$. 

\paragraph{Computing the grand potentials.}
We highlight a subtle point regarding the PBM implementation in the VASPsol package~\cite{2019_JChemPhys_Mathew_Hennig}, which we used in the present study. As the VASPsol authors emphasize~\cite{2019_JChemPhys_Mathew_Hennig}, an energy contribution $\Delta E = e\,\phi_{\mathrm{elyte}} N_e$ must be added to the free energy $A_{\mathrm{VASP}}$ printed by the VASP core package in order to obtain the total free energy $A = A_{\mathrm{VASP}} + \Delta E$ of the system. Furthermore, in the Kohn-Sham-Mermin DFT approach, the electron chemical potential $\mu_e = E_{\mathrm{F}}$ is equal to the Fermi energy of the Fermi-Dirac distribution on the Kohn-Sham eigenvalues, which is directly printed by the VASP core package. Finally, the chemical part $\mu_i^0$ of the ion chemical potentials is simply set to zero. Therefore, the PBM grand potential according to Eq.~\eqref{eq_grand_pot_PBM_combined} is obtained from the VASPsol implementation as 
\begin{align}
\Omega_{\mathrm{PBM}}\, & =\, A_{\mathrm{VASP}} + e\,\phi_{\mathrm{elyte}} N_e - (\mu_e + e\,\phi_{\mathrm{elyte}}) N_e \nonumber \\
& =\, A_{\mathrm{VASP}} - E_{\mathrm{F}} N_e 
\label{eq_grand_pot_PBM_VASP}
\end{align}
To compute the HBM grand potential according to Eq.~\eqref{eq_grand_pot_HBM}, we note that the mean electrostatic potential is set to zero within VASP, $\bar{\phi} = 0$ , so
\begin{align}
\Omega_{\mathrm{HBM}} \,=\, A_{\mathrm{VASP}} - E_{\mathrm{F}} N_e
\label{eq_grand_pot_HBM_VASP}
\end{align} 
and we obtain the same form as in Eq.~\eqref{eq_grand_pot_PBM_VASP}. We provide our processing scripts for the PBM and HBM calculations with VASPsol in an online repository~\cite{scripts_repo}.

\section{Results and Discussion}

\begin{figure}[t]
\centering
\includegraphics[scale=0.27]{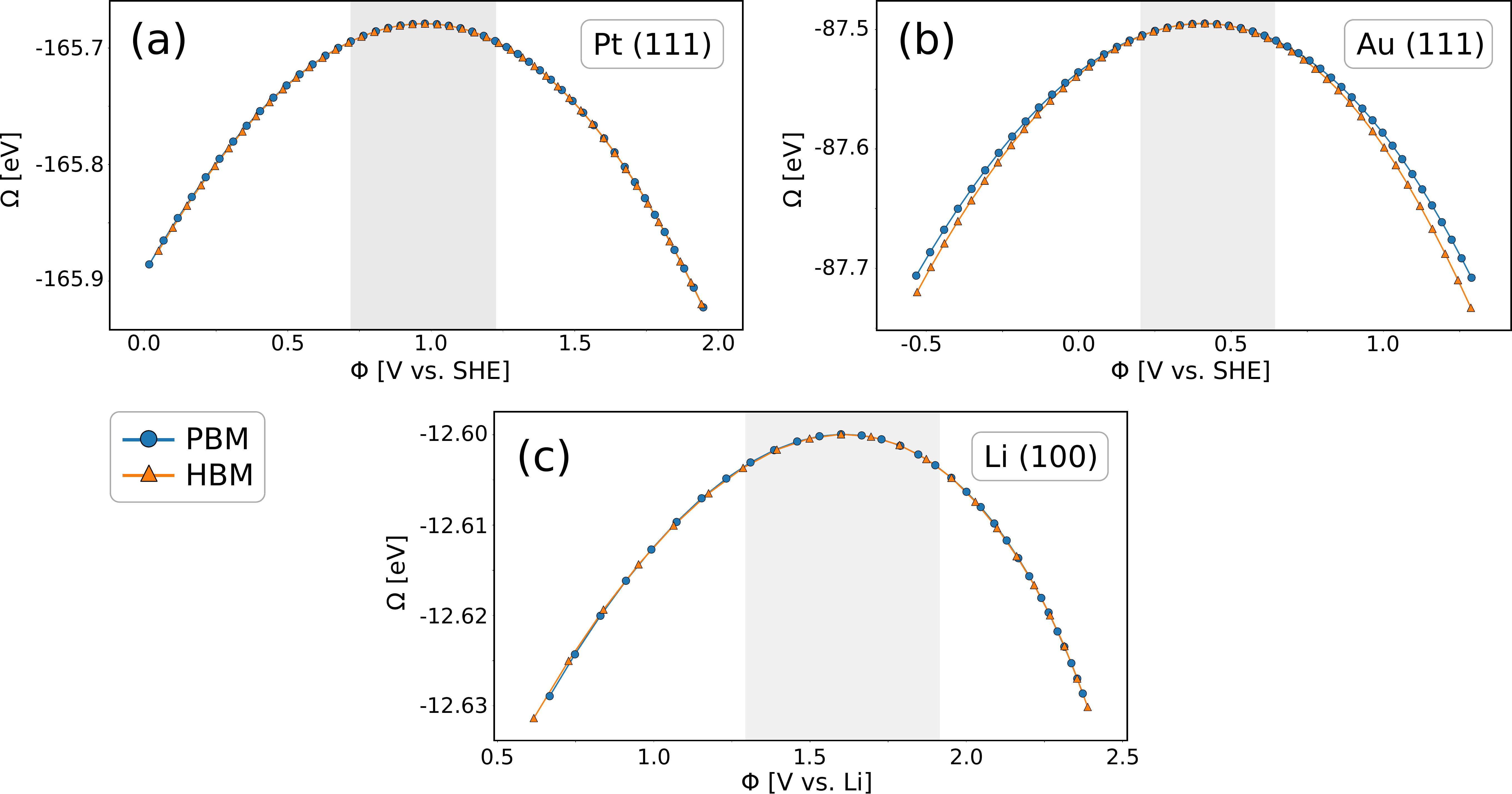}
	\caption{Comparison of the HBM vs. PBM grand potentials $\Omega$ as a function of the electrode potential, computed for (a) Pt(111), (b) Au(111) and (c) Li(100) electrodes in contact with an implicit solvent with $\epsilon_b$ = 78.4 for Pt and Au and $\epsilon_b$ = 89.9 for Li, corresponding to water and ethylene carbonate, respectively. Grey regions highlight the potential ranges (PZC $\pm$ 0.25 V) on which the fitted values for the capacitances in Table~\ref{tab_capacitance} have been computed.}
\label{fig_hbm_vs_debye}
\end{figure}

Figure~\ref{fig_hbm_vs_debye} shows a comparison between the HBM grand potential, Eq.~\eqref{eq_grand_pot_HBM}, and the PBM grand potential, Eq.~\eqref{eq_grand_pot_PBM_combined}, as a function of the electrode potential computed for Pt(111), Au(111) and Li(100) electrodes in contact with implicit electrolyte. It is obvious that the grand potentials obtained from the two different counter-charge models are almost identical, despite the ``unphysical'' extension of the homogeneous background charge across the bulk of the electrode slab. This is a remarkable finding, because the definition of the HBM grand potential according to Eq.~\eqref{eq_grand_pot_HBM} neither involves any homogeneous background corrections, nor any specification of the actual surface electron number. Table~\ref{tab_capacitance} presents the fitted values of the capacitances defined in Eqs.~\eqref{eq_capacitance_Phi_N_HBM} and \eqref{eq_capacitance_Omega_Phi_HBM} for HBM and PBM at the potential of zero charge (PZC), indicated by the ``$0$'' superscript. As discussed in the Theory and Methods section above, the capacitance definitions are precisely equivalent in the PBM, which is confirmed by their fitted values for all systems. In the HBM, however, they correspond to different quantities, resulting in entirely different fitted values. Interestingly, the fitted values of the capacitance $C_{\Omega \Phi}^0$ obtained for the HBM agree within less than 10\% with those obtained for the PBM. Since $C_{\Omega \Phi}^0$ corresponds to the curvature of the $\Omega(\Phi)$ curve, this agreement is directly correlated with the agreement of the $\Omega(\Phi)$ curves shown in Figure~\eqref{fig_hbm_vs_debye}. Thus, the grand potentials $\Omega(\Phi)$ and the capacitances $C_{\Omega \Phi}^0$ at the PZC, agree remarkably well between the HBM and PBM, despite their very different counter-charge distributions. Noting that, within the PBM, $C_{\Omega \Phi}$ is precisely equal to the interface capacitance of Eq.~\eqref{eq_capacitance_Phi_q_PBM}, we conclude that $C_{\Omega \Phi}^0$ for the HBM provides a satisfactory representation of the ``true'' interface capacitance at the PZC, at least within the common limitations of HBM and PBM to be discussed below.\footnote{We emphasize that, in the present context, we understand the ``true'' interface capacitance as the one defined by Eq.~\eqref{eq_capacitance_Phi_q_PBM} \emph{for the given model} where $q$ refers to the interfacial charge, only. The question whether the value of this ``true'' model interface capacitance agrees with the real interface capacitance of the real physical system is a different one and lies outside of the scope of the present work, where we exclusively focus on a comparison between the computational HBM and PBM methods.} We furthermore see that the straightforward definition of the HBM grand potential according to Eqs.~\eqref{eq_grand_pot_HBM} and \eqref{eq_grand_pot_HBM_VASP}, which avoids any \textit{a posteriori} energy corrections, provides results in excellent agreement with the PBM. This finding will be further rationalized in the following.

\begin{table}[tb]
\centering
\begin{tabular}{|c|c|c|c|c|}
\hline
	\multirow{2}{*}{Electrode} & \multicolumn{2}{|c|}{PBM} & \multicolumn{2}{|c|}{HBM} \\ \cline{2-5}
 & $C_{\Phi N}^0$      & $C_{\Omega \Phi}^0$  &  $C_{\Phi N}^0$     & $C_{\Omega \Phi}^0$ \\ \hline
	Li(100)           & 5.12        & 5.07     & 14.13       & 5.21  \\
	Pt(111)           & 13.98       & 13.98    & 33.60       & 14.05 \\
	Au(111)           & 13.71       & 13.59    & 32.84       & 14.62 \\ \hline
\end{tabular}
\caption{Comparison between the PBM and HBM methods of the fitted values for the capacitances $C_{\Phi N}^0$ and $C_{\Omega \Phi}^0$ at the PZC obtained from the computed $\Phi(N_e)$ and $\Omega(\Phi)$ curves, respectively, for Li(100), Pt(111), and Au(111) electrodes in contact with implicit electrolyte. Capacitance values are given in $\mathrm{\mu F\,cm^{-2}}$.}
\label{tab_capacitance}
\end{table}

\begin{figure}[t]
\centering
\includegraphics[scale=0.85]{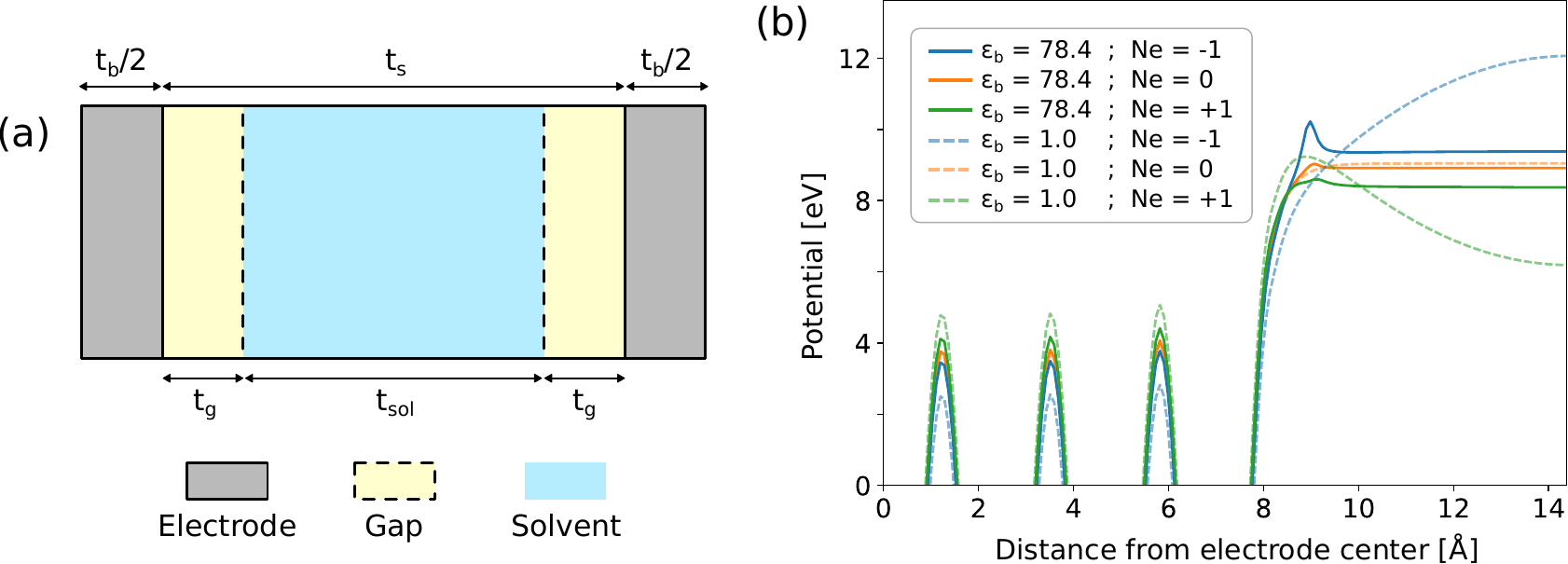}
	\caption{(a) Scheme of an electrode slab simulation cell with electrode (grey), gap (yellow) and solvent (blue) regions. (b) Potential (averaged over $xy$-area) along the $z$-axis for a Pt(111) electrode in contact with an implicit solvent with $\epsilon_b = 78.4$ and $\epsilon_b = 1.0$ at different charge states $N_e = -1, 0, +1$ using the HBM.}
\label{fig_scheme_models}
\end{figure}

\subsection{Separating bulk electrode and electrolyte contributions in HBM}

To explore the reason why the capacitance $C_{\Omega \Phi}^0$ in the HBM provides a good description of the interface capacitance at the PZC, we investigate a simplified model where the HBM contributions within the bulk electrode and the electrolyte regions are treated separately. A schematic simulation cell of a periodic electrode-slab model is shown in Figure~\ref{fig_scheme_models}(a). Here, $t_b$ denotes the total electrode slab thickness ($b$: \textit{bulk}) and $t_s$ corresponds to the width of the remaining interspace region ($s$: \textit{space} or \textit{surface}), which contains the implicit solvent/electrolyte. A sub-partitioning of $t_s$ into an actual solvent-region width $t_{sol}$ and twice a gap width $t_g$ will be treated later on. For now, we continue with the coarse partitioning into $t_b$ and $t_s$. Accordingly, the total width of the simulation cell in the $z$-direction perpendicular to the slab surface is $T = t_b + t_s$, and the volume fractions of the bulk-electrode and interspace regions are $\alpha_b = t_b/T$ and $\alpha_s = t_s/T$, respectively.

Imagine that we could independently change the homogeneous background charge within both the bulk electrode and the electrolyte regions (while, at the same time, changing the electron number accordingly to preserve charge neutrality). We denote the respective excess electron numbers by $N_b$ and $N_s$, and the Helmholtz free energy $A(N_b, N_s)$ becomes a function of both. We next perform a second-order expansion around the state with zero excess electrons,
\begin{align} 
\label{eq_A_tot_expanded}
A(N_b, N_s) \approx A^0 + \mu^0_b N_b + \mu^0_s N_s + \frac{1}{2}\,\eta^0_b N_b^2 + \frac{1}{2}\,\eta^0_s N_s^2 + \eta^0_{bs} N_b N_s
\end{align}
and we neglect higher-order terms, which do not contribute to the capacitance at the potential of zero charge, see Eq.~\eqref{eq_capacitance_Omega_Phi_HBM}. Here, 
\begin{align}
\mu_{\sigma} = \frac{\partial A}{\partial N_{\sigma}}
\label{eq_chem_pot_model}
\end{align}
and 
\begin{align}
\eta_{\sigma} = \frac{\partial^2 A}{\partial N_{\sigma}^2}
\label{eq_chem_hard_model}
\end{align}
correspond to the chemical potential and chemical hardness, respectively, of the bulk ($\sigma=b$) and surface/interspace ($\sigma=s$) sub-systems, and the ``$0$'' superscript denotes the state with zero excess electrons. Because the derivatives are defined by charge-neutral variations in both electron number and homogeneous background (within the respective region), we note that the chemical potentials $\mu_{\sigma} = \mu_{e,\sigma} + \mu_{bg,\sigma}$ are the sum of the electron and background chemical potentials, where $\mu_{bg,\sigma} = e\bar{\phi}_{\sigma}$ is simply determined by the average value of the electrostatic potential $\bar{\phi}_{\sigma}$ within each of the regions~\cite{2020_PCCP_Kopac}. The last term in Eq.~\eqref{eq_A_tot_expanded} with
\begin{align}
\eta_{bs} = \frac{\partial^2 A}{\partial N_b \partial N_s}
\label{eq_interactions_model}
\end{align}
quantifies the interactions between the bulk/surface excess-electron and background charges. 

Using the HBM, however, we cannot control $N_b$ and $N_s$ individually, but we fix the total excess electron number $N_e = N_b + N_s$, compensated by homogeneous background charge across the entire simulation cell. The fraction of homogeneous background charge in the bulk electrode region is equal to the respective volume fraction $\alpha_b$. Because we only consider metallic electrodes, we assume that a corresponding fraction of $N_b = \alpha_b N_e$ excess electrons accumulates within the bulk electrode region to screen the homogeneous background charge therein~\cite{2020_PCCP_Kopac}. Accordingly, the fraction of excess electrons accumulated at the electrode surface is given by $N_s = \alpha_s N_e$, corresponding to the homogeneous background charge in the electrolyte region (note that $\alpha_s + \alpha_b = 1$). 

In order to compute the HBM grand potential $\Omega = A - (\mu_e + e\bar{\phi})N_e$ according to Eq.~\eqref{eq_grand_pot_HBM}, we use the free energy expansion of Eq.~\eqref{eq_A_tot_expanded} and we also expand the second term $(\mu_e + e\bar{\phi})N_e$. The mean electrostatic potential $\bar{\phi} = \alpha_b\bar{\phi}_b + \alpha_s\bar{\phi}_s$ of the entire simulation cell can be written as the weighted average of the mean values $\bar{\phi}_b$ and $\bar{\phi}_s$ across the bulk and interspace regions, respectively. Since the entire simulation cell is electronically equilibrated, the electron chemical potentials of the two regions are equal, $\mu_{e,b} = \mu_{e,s} = \mu_e$. We thus obtain
\begin{align}
\label{eq_omega_term_split}
(\mu_e + e\bar{\phi})N_e & = \mu_e N_e + e\bar{\phi}_b\alpha_b N_e + e\bar{\phi}_s\alpha_s N_e \nonumber\\
& = \mu_e (N_b + N_s) + e\bar{\phi}_b N_b + e\bar{\phi}_s N_s \nonumber\\
& = \mu_b N_b + \mu_s N_s \nonumber\\
& \approx \mu^0_bN_b + \eta^0_bN_b^2 + \mu^0_sN_s + \eta^0_sN_s^2 + 2\eta_{bs}^0 N_b N_s
\end{align}
where we used $N_b = \alpha_b N_e$, $N_s = \alpha_s N_e$, and $N_e = N_b + N_s$. In the last step, the bulk and surface chemical potentials were again expanded around the zero-excess state, $\mu_b \approx \mu_b^0 + \eta_b^0 N_b + \eta_{bs}^0 N_s$ and $\mu_s \approx \mu_s^0 + \eta_s^0 N_s + \eta_{bs}^0 N_b$, respectively, which follows from Eqs.~\eqref{eq_chem_pot_model}, \eqref{eq_chem_hard_model}, and \eqref{eq_interactions_model}. Using Eqs.~\eqref{eq_A_tot_expanded} and \eqref{eq_omega_term_split}, the HBM grand potential reads to second order
\begin{equation}
\label{eq_grand_pot_HBM_model}
\Omega \approx A^0 - \frac{1}{2}\left(\alpha_b^2\eta^0_b + \alpha_s^2\eta^0_s + 2\alpha_b \alpha_s\eta^0_{bs}\right)N_e^2
\end{equation}

To define the electrode potential, we introduce a further approximation, restricting our model to \emph{high-dielectric-constant} solvents with $\epsilon_r \gg 1$, e.g. water with $\epsilon_r \approx 80$. Then, the electrostatic potential in the electrolyte region assumes an approximately constant plateau value $\phi_{\mathrm{elyte}}$, and the curvature due to the homogeneous background charge can be neglected, see Figure~\ref{fig_scheme_models}(b). Consequently, the mean value $\bar{\phi}_s$ across the interspace region becomes approximately equal to the plateau value $\phi_{\mathrm{elyte}}$. The electrode potential, defined according to Eq.~\eqref{eq_electrode_potential}, is then simply given by 
\begin{align}
-e\,\Phi \,\approx\, \mu_{e,s} + e\bar{\phi}_s \,=\, \mu_s
\label{eq_electrode_potential_HBM_model}
\end{align}
with the chemical potential $\mu_s = \mu_{e,s} + \mu_{bg,s}$, as discussed above. We develop the electrode potential to first order around the potential of zero charge (PZC) $\Phi^0$ as $\Phi \approx \Phi^0 - (\alpha_s\eta_s^0+\alpha_b\eta_{bs}^0) N_e/e$, which follows from inserting the first-order expansion of $\mu_s$, see above, into Eq.~\eqref{eq_electrode_potential_HBM_model}. Resolving for $N_e$ and inserting into Eq.~\eqref{eq_grand_pot_HBM_model} leads to the grand potential as a function of the electrode potential up to second order,
\begin{align}
\label{eq_grand_pot_Phi_HBM_model}
\Omega \approx A^0 - \frac{1}{2}\,\frac{\alpha_b^2\eta^0_b + \alpha_s^2\eta^0_s + 2\alpha_b \alpha_s\eta^0_{bs}}{\left(\alpha_s\eta_s^0+\alpha_b\eta_{bs}^0\right)^2}\,e^2 (\Phi-\Phi^0)^2
\end{align}

We now assess the capacitance definitions for the present model. Firstly, according to Eq.~\eqref{eq_capacitance_Phi_q_PBM}, we \emph{define} the ``true'' interface capacitance as
\begin{align}
\frac{1}{C_s} \,=\, \frac{\partial \Phi}{\partial q_s} \,=\, \frac{1}{e^2}\,\frac{\partial \mu_s}{\partial N_s} \,=\, \frac{\eta_s}{e^2}
\label{eq_C_surface_HBM_model}
\end{align}
where $q_s = -e N_s$ is the excess charge at the electrode surface, and we used Eqs.~\eqref{eq_electrode_potential_HBM_model}, \eqref{eq_chem_pot_model}, and \eqref{eq_chem_hard_model}. Note that the surface excess electron number $N_s$ cannot be directly specified as an input parameter of HBM-DFT calculations. Instead, the total excess electron number $N_e$ of the simulation cell is fixed as an input, which corresponds to the capacitance $C_{\Phi N}$ defined in Eq.~\eqref{eq_capacitance_Phi_N_HBM}. Using $N_s = \alpha_s N_e$, we find
\begin{align}
\frac{1}{C_{\Phi N}} \,=\, \frac{\alpha_s}{C_s} \,=\, \frac{\alpha_s\eta_s}{e^2}
\label{eq_C_Phi_N_HBM_model}
\end{align}
The capacitance $C_{\Omega \Phi}^0$, defined by Eq.~\eqref{eq_capacitance_Omega_Phi_HBM}, at the PZC is obtained from Eq.~\eqref{eq_grand_pot_Phi_HBM_model},
\begin{align}
C_{\Omega \Phi}^0 \,=\, C_s^0\ \frac{1 + \left[(\alpha_b^2\eta^0_b+2\alpha_b \alpha_s\eta^0_{bs})/(\alpha_s^2\eta^0_s)\right]}{\left(1+\left[\alpha_b\eta_{bs}^0/(\alpha_s\eta_s^0)\right]\right)^2}
\label{eq_C_Omega_Phi_HBM_model}
\end{align}
where we used Eq.~\eqref{eq_C_surface_HBM_model} for the ``true'' interface capacitance $C_s^0$ at the PZC. 

For metal electrodes, the delocalized electrons effectively screen the homogeneous background charge in the bulk electrode region, resulting in a large ``bulk capacitance'', or small bulk chemical hardness $\eta^0_b$ in comparison to $\eta^0_s$ of the surface, so $\eta^0_b \ll \eta^0_s$. This is, for instance, verified for the Pt(111) slab, for which we obtain $\eta^0_b = 0.052\,\mathrm{eV} \ll \eta^0_s \approx 2.1\,\mathrm{eV}$.\footnote{Here, $\eta^0_s$ was computed from $C_{\Omega \Phi}^0$ using the approximation $C_s^0 \approx C_{\Omega \Phi}^0$. For the bulk chemical hardness $\eta^0_b$, an independent HBM calculation of the corresponding Pt-bulk cell was performed, as described in the Computational Details section, and $\eta^0_b$ was obtained from a second-order polynomial fit of the resulting $A_b(N_b)$ curve.} The metallic screening makes the bulk electrode region charge neutral, and, due to overall charge neutrality, the same must hold for the surface/interspace part. Therefore, the bulk--surface interactions quantified by $\eta^0_{bs}$ result from higher-order multipole interactions between both regions, which are expected to be significantly weaker than the intra-region interactions quantified by $\eta^0_b$ and $\eta^0_s$, so $\eta^0_{bs} \ll \eta^0_b$ and $\eta^0_{bs} \ll \eta^0_s$, which is confirmed by our computational results discussed below. 

Under these conditions, the terms in square brackets in Eq.~\eqref{eq_C_Omega_Phi_HBM_model} are negligible and it follows that $C_{\Omega \Phi}^0 \approx C_s^0$. Thus, for metal electrodes, the HBM capacitance $C_{\Omega \Phi}^0$ at the PZC is in good agreement with the ``true'' HBM interface capacitance $C_s^0$. This provides a first part of the explanation for the good agreement between the HBM $C_{\Omega \Phi}^0$ and the PBM capacitance, as established by the results in Table~\ref{tab_capacitance}. The second part of the explanation, i.e. why $C_s^0$ of the HBM, in turn, is approximately equal to the PBM capacitance, will be discussed further below.

\subsection{Advancement of the Homogeneous Background Method}

\begin{figure}[t]
\centering
\includegraphics[scale=0.26]{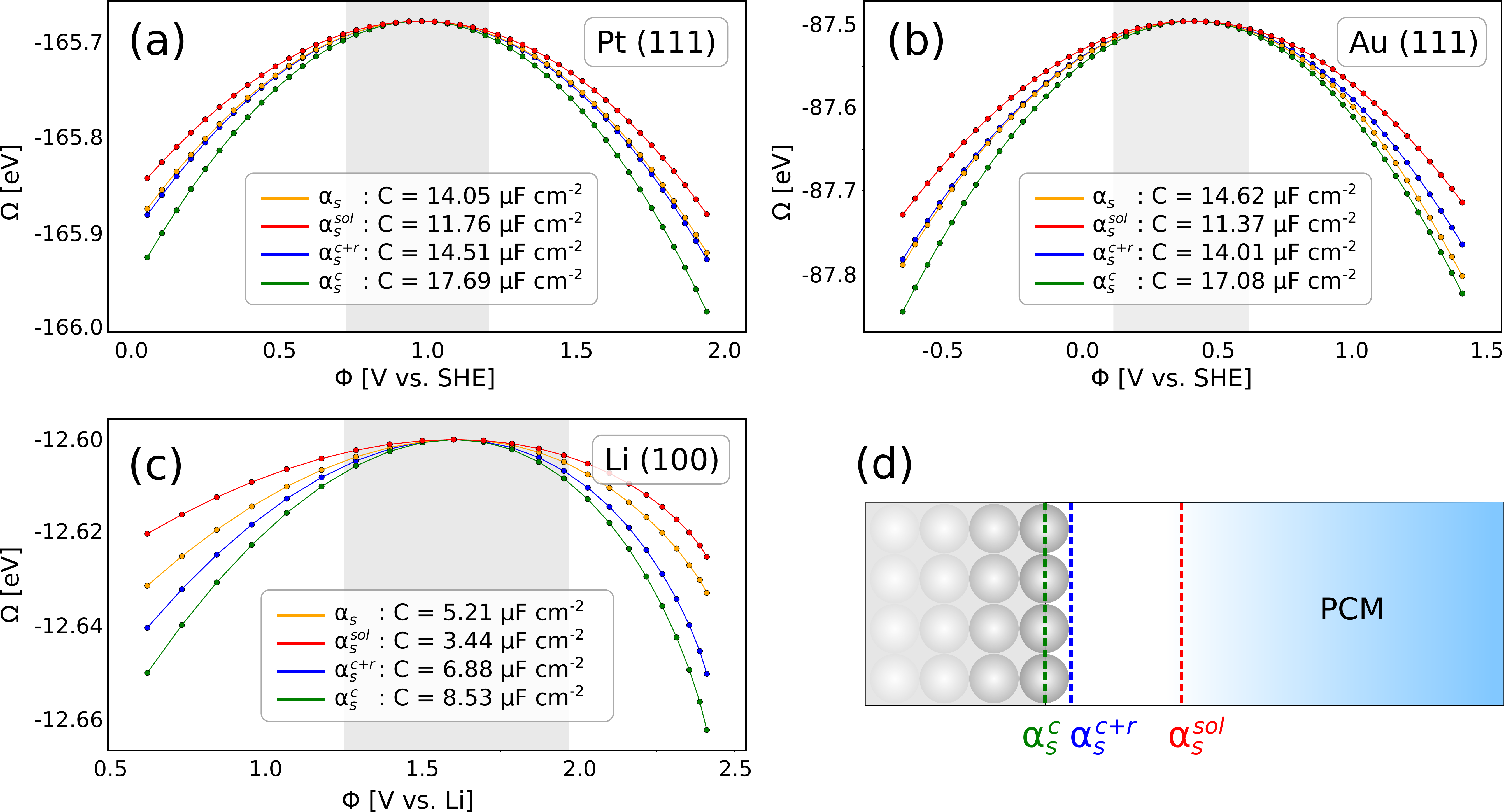}
	\caption{Comparison between the advanced HBM version and the original HBM approach: Grand potential $\Omega$ as a function of the electrode potential $\Phi$ computed for (a) Pt(111), (b) Au(111), and (c) Li(100) in contact with implicit electrolyte. The results of the original HBM are shown for the three different definitions of $\alpha_s$, see text. (d) Scheme of the interface between an electrode slab and (implicit) electrolyte with three possible definitions of the electrode--electrolyte boundary, and thus of $\alpha_s$, for the original HBM approach.}
\label{fig_hbm_with_without}
\end{figure}

Our finding $C_{\Omega \Phi}^0 \approx C_s^0$ enables an advancement of the HBM, both facilitating  and extending the applicability of the method. The advanced version is simply represented by the calculation of the grand potential $\Omega$ according to Eqs.~\eqref{eq_grand_pot_HBM} or \eqref{eq_grand_pot_HBM_VASP}, which can be readily performed with the knowledge of the total electron excess $N_e$ of the simulation cell that is specified as an input parameter for the HBM-DFT calculations. The interface capacitance can then be derived in good approximation from the curvature of the computed $\Omega(\Phi)$ curve, see Eq.~\eqref{eq_capacitance_Omega_Phi_HBM}. Unlike the original approach~\cite{2006_AngewandteChem_Filhol, 2020_PCCP_Kopac}, the calculation of the $\Omega(\Phi)$ curve according to the advanced HBM does not require an estimation of the number $N_s$ of excess electrons at the electrode surface. The consistency between the advanced and original approaches is discussed in the following section and demonstrated by the results in Figure~\ref{fig_hbm_with_without}. 

The meaningfulness of the obtained $\Omega(\Phi)$ curve is further demonstrated by the good agreement with the $\Omega(\Phi)$ curve from PBM, shown in Figure~\ref{fig_hbm_vs_debye} and discussed above. We can now explain this agreement. At the PZC, there is no counter charge present, so the HBM and PBM grand potentials are identical. Furthermore, the grand potential has a maximum at the PZC, so the first derivative vanishes, cf. Eq.~\eqref{eq_deriv1_Omega_mu_PBM} with $N_e = 0$. Finally, as shown below, the ``true'' interface capacitance $C_s$ of the HBM is approximately equal to the capacitance of the PBM. Therefore, our result $C_{\Omega \Phi}^0 \approx C_s^0$ means that also the curvature of the $\Omega(\Phi)$ curve at the PZC is approximately equal in HBM and PBM, cf. Eqs.~\eqref{eq_capacitance_Omega_Phi_PBM} and \eqref{eq_capacitance_Omega_Phi_HBM}. At least up to second order in $\Phi$ around the PZC, a good agreement is thus established between the $\Omega(\Phi)$ curves of the advanced HBM and PBM. 

\begin{table}[tb]
\centering
\begin{tabular}{|c|c|c|c|c|}
\hline
	System    & $\alpha_s^{c}$ & $\alpha_s^{c+r}$ & $\alpha_s^{sol}$ & $\alpha_s = \frac{C_{\Omega \Phi}^0}{C_{\Phi N}^0}$ \\ \hline
	Li(100)       &  0.59 & 0.48  & 0.24   &  0.37   \\
	Pt(111)       &  0.52 & 0.43  &  0.35  &  0.42   \\
	Au(111)       &  0.51 & 0.42  & 0.34   & 0.44    \\
\hline
\end{tabular}
\caption{Comparison of the different definitions for $\alpha_s$, see text, for Li(100), Pt(111), and Au(111) electrodes in contact with implicit electrolyte. The last column contains the ``intrinsic'' estimate of $\alpha_s$ provided by the advanced HBM according to Eq.~\eqref{eq_active_fraction}.}
\label{tab_alpha}
\end{table}

\subsection{Intrinsic estimation of ``active'' electron fraction}

The approximation $C_{\Omega \Phi}^0 \approx C_s^0$ of Eq.~\eqref{eq_C_Omega_Phi_HBM_model}, combined with Eq.~\eqref{eq_C_Phi_N_HBM_model}, provides an estimation of the ``active'' fraction $N_s/N_e = \alpha_s$ of excess electrons accumulated at the electrode--electrolyte interface,
\begin{align} 
\label{eq_active_fraction}
\alpha_s = \frac{C_s^0}{C_{\Phi N}^0} \approx \frac{C_{\Omega \Phi}^0}{C_{\Phi N}^0} 
\end{align}
without requiring any \textit{a priori} specification of a boundary between the bulk electrode and interspace regions. 

In contrast, the original HBM approach~\cite{2006_AngewandteChem_Filhol, 2020_PCCP_Kopac} relies on an \textit{a priori} definition of the interspace volume fraction $\alpha_s$. In fact, several reasonable choices for $\alpha_s$ exist, according to the definition of the boundary between electrode and interspace region, as visualized in Figure~\ref{fig_hbm_with_without}(d). First, the location of the center of the electrode-slab surface atoms can be used to define the boundary, yielding an interspace volume fraction $\alpha_s^{c}$. Second, the atomic radius can be added to the center location of the surface atoms, resulting in $\alpha_s^{c+r}$. And third, the interspace region can be defined simply as the implicit solvent region of VASPsol, leading to $\alpha_s^{sol}$. As shown in Table~\ref{tab_alpha}, these definitions follow the order $\alpha_s^{sol} < \alpha_s^{c+r} < \alpha_s^{c}$, with differences in value up to a factor of two for the given (typical) simulation cell dimension. Interestingly, the ``intrinsic'' estimate of $\alpha_s$ provided by the advanced HBM according to Eq.~\eqref{eq_active_fraction} lies in the middle of the value range and agrees rather well with the definition $\alpha_s^{c+r}$. Likewise, as shown in Figure~\ref{fig_hbm_with_without}(a)--(c), the $\Omega(\Phi)$ curve directly computed according to Eq.~\eqref{eq_grand_pot_HBM} without \textit{a posteriori} corrections falls in between the $\Omega(\Phi)$ curves computed according to the original HBM approach using the three definitions of $\alpha_s$. These results demonstrate that the advanced HBM provides a method that is not only consistent with the original HBM approach, but also avoids the problem of defining $\alpha_s$, by providing an intrinsic ``detection'' of the ``active'' fraction of excess electrons. This will be highly useful for the application of the HBM in electrochemical-interface simulations with mixed explicit--implicit or fully explicit solvent models and in presence of surface adsorbate species that further complicate an \textit{a priori} definition of the electrode--electrolyte boundary, and thus of $\alpha_s$.

\begin{figure}[t]
\centering
\includegraphics[scale=0.35]{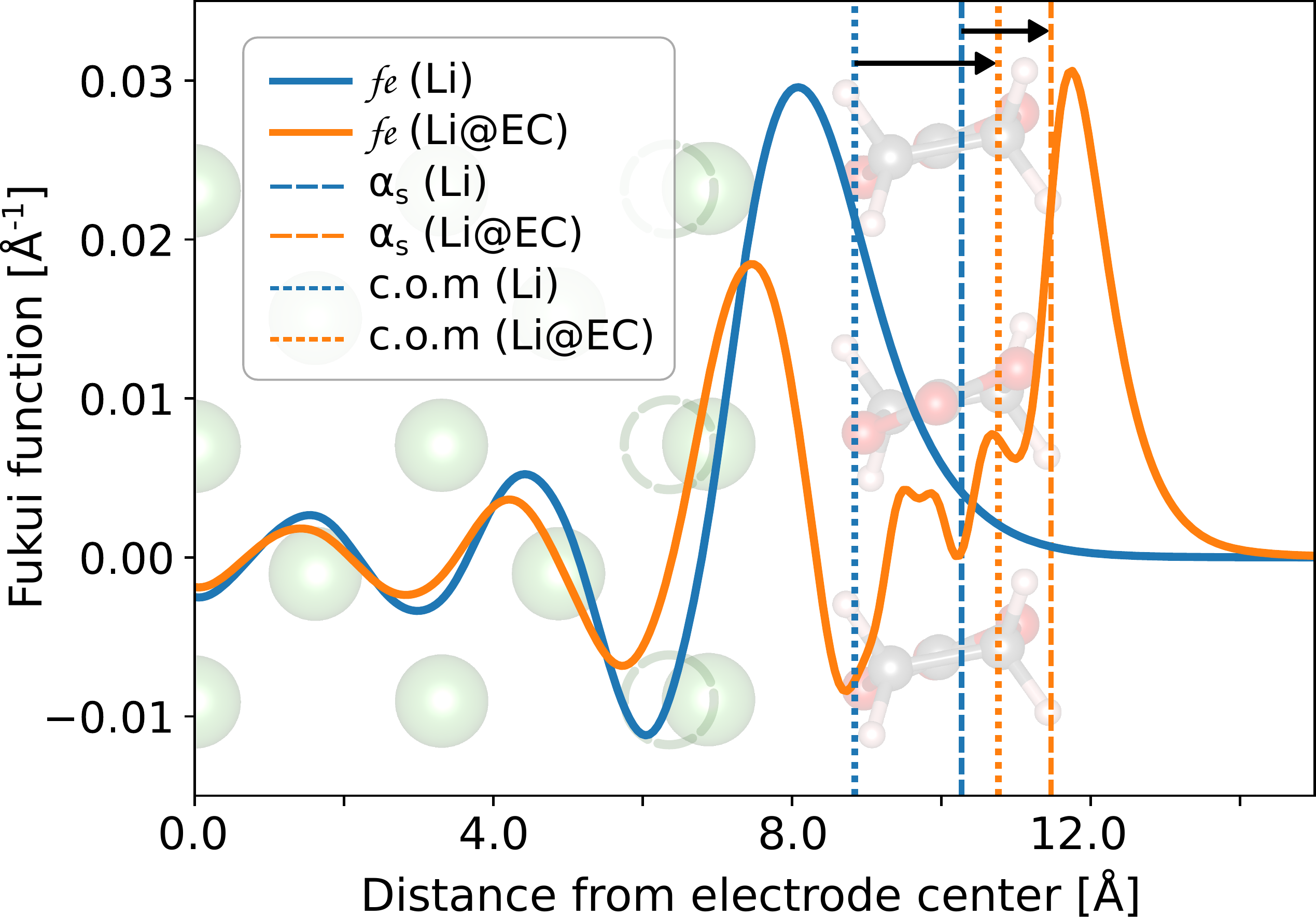}
\caption{Effective location of the electrode--electrolyte boundary (dashed vertical lines) within the HBM, computed from the $\alpha_s$ values for the Li(100) and Li(100)@EC systems without and with an explicit layer of EC solvent molecules, respectively. The normalized Fukui functions from the PBM, averaged over the surface-parallel $xy$-coordinates of the two systems, are shown for comparison, with the dotted vertical lines indicating the center of mass (c.o.m) of the Fukui functions. The shifts in the effective boundary locations due to the explicit EC molecules are indicated by arrows. The relaxed Li(100)@EC structure is shown in the background for orientation. The Li surface atoms of the relaxed Li(100) structure are indicated by dashed circles.}
\label{fig_fukui}
\end{figure}

To demonstrate this aspect, we performed HBM-DFT calculations for the Li(100) surface, where the first layer of ethylene carbonate (EC) solvent molecules was explicitly modelled, combined with implicit EC solvent in the remaining interspace region. To accommodate the explicit EC molecules, the simulation cell comprised a 2x2 surface cell of the Li(100) slab (with 9 atomic layers) and a $40\,\mathrm{\AA}$-wide interspace between the periodic slab images. For the Li(100) and Li(100)@EC systems without and with the explicit EC layer, respectively, Eq.~\eqref{eq_active_fraction} yielded $\alpha_s$ values of $0.622$ and $0.578$, respectively. Since $\alpha_s$ corresponds to the volume fraction of the interspace region, the ``effective'' location of the boundary between the electrode and interspace (electrolyte) can be readily calculated. These are shown as dashed vertical lines in Figure~\ref{fig_fukui}, together with the normalized electronic Fukui function $f_e(z)$, averaged over the surface-parallel $xy$-coordinates, as a function of the surface-perpendicular $z$-coordinate. The electronic Fukui function $f_e(\mathbf{r}) = \partial n_e(\mathbf{r})/\partial N_e$ is the derivative of the electron density $n_e(\mathbf{r})$ with respect to changes in electron number $N_e$, and it represents the distribution of the excess electrons at the electrode--electrolyte interface~\cite{2014_JPhysChemC_Filhol_Doublet}. The Fukui center of mass along the surface normal provides a natural definition of the electrode surface location for a dipole analysis of the interface capacitance~\cite{2021_PhysRevB_Binninger}. The Fukui function shown in Figure~\ref{fig_fukui} has been computed using the PBM, in order to avoid the contribution of the bulk excess electrons within the HBM that would bias the Fukui center of mass. In presence of the explicit EC molecule layer, the Fukui function gets split into two dominant peaks, one centered at the surface Li atoms, and the other at the EC molecules. This means that excess electrons spill over onto the latter, and the EC molecules participate in the electronic charging of the interface~\cite{hagopian_ab_2021}. Consequently, this first layer of EC solvent cannot be uniquely ascribed to either the electrode or the electrolyte side of the interface, making an \textit{a priori} definition of the electrode--electrolyte boundary difficult. The intrinsic detection of the boundary location according to the $\alpha_s$ values, however, correctly reflects the outwards shift of the Fukui function in presence of the explicit EC solvent layer in comparison with the purely implicitly solvated Li(100) surface. Even quantitatively, a consistent behavior is found: Whereas the boundary computed from $\alpha_s$ is shifted by $1.2\,\mathrm{\AA}$, the center of mass (c.o.m) of the Fukui function in the $z$-direction is shifted by $1.9\,\mathrm{\AA}$. Given the fact that both represent different definitions of the boundary location, the quantitative agreement between these values is satisfying and corroborates the validity of the approximation in Eq.~\eqref{eq_active_fraction}.

\subsection{Comparison between ``true'' interface capacitance of HBM and PBM}

We finally return to the question for the origin of the good agreement between the HBM and PBM grand potentials defined in Eqs.~\eqref{eq_grand_pot_HBM} and \eqref{eq_grand_pot_PBM_combined}, respectively, as shown in Figure~\ref{fig_hbm_vs_debye}. As discussed before, this is directly related to the agreement between the HBM and PBM capacitances $C_{\Omega \Phi}$, presented in Table~\ref{tab_capacitance}. Within the PBM, $C_{\Omega \Phi}$ of Eq.~\eqref{eq_capacitance_Omega_Phi_PBM} is precisely equal to the interface capacitance $C$ of Eq.~\eqref{eq_capacitance_Phi_q_PBM}. We further demonstrated, cf. Eq.~\eqref{eq_C_Omega_Phi_HBM_model}, that the HBM capacitance $C_{\Omega \Phi}$ is approximately equal to the ``true'' HBM interface capacitance $C_{s}$. Therefore, the ``true'' interface capacitances of HBM and PBM seem to be approximately equal, which appears surprising given the very different counter-charge distributions in the electrolyte region for the two models.

This behavior can be understood by taking into account the existence of a ``gap'' between the electrode surface and the beginning of the implicit solvent region~\cite{2012_PhysRevB_Letchworth-Weaver_Arias, 2015_JChemTheoComp_Filhol, 2019_JChemPhys_Hoermann_Marzari}, see Figure~\ref{fig_scheme_models}(a). In the present context, we consider the ``gap'' simply as a numerical aspect of the implicit solvent model, and we refer to other literature for a discussion of its relation to the Helmholtz capacitance of the physical system~\cite{2012_PhysRevB_Letchworth-Weaver_Arias}. The width $t_g$ of this gap is numerically controlled within the VASPsol implementation by the parameter $n_c$ that represents the critical value of the electron density defining the location of the implicit solvent boundary~\cite{2019_JChemPhys_Mathew_Hennig}. Similar parameters are also defined in other implementations of implicit solvent models for DFT codes~\cite{2019_JChemPhys_Hoermann_Marzari}. Since the gap region does not contain implicit solvent, its dielectric properties only result from the small residual electronic density within this region, corresponding to a significantly reduced effective dielectric constant $\epsilon_{r,g}$ and a \textit{gap capacitance}~\cite{2012_PhysRevB_Letchworth-Weaver_Arias} $C_{g} = (2\mathcal{A})(\epsilon_0\epsilon_{r,g})/t_g$, where we take into account \emph{both} electrode--electrolyte interfaces of the simulation cell with a total area of $2\mathcal{A}$.

\begin{figure}[t]
\centering
\includegraphics[scale=0.5]{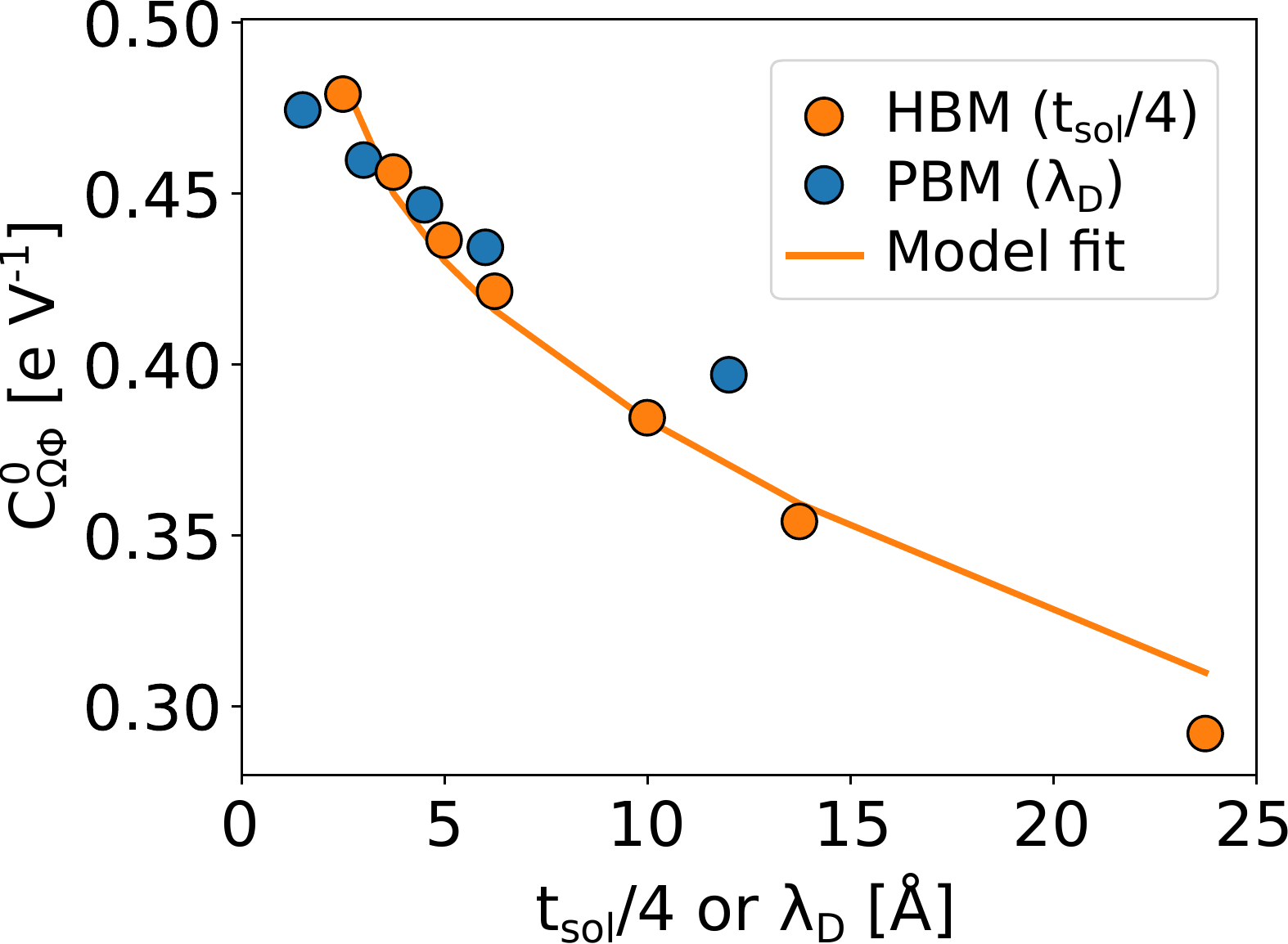}
\caption{HBM and PBM capacitance $C_{\Omega \Phi}^0$ at the PZC computed for the Pt(111) slab in contact with implicit water solvent as a function of $t_{sol}/4$ and $\lambda_{\mathrm{D}}$, respectively, together with the fitted curve according to the electrostatic model for the HBM, see Supporting Information.}
\label{fig_cap_vs_ts}
\end{figure}

For the PBM, the overall interface capacitance $C$ is given by the series of the gap capacitance  $C_g$ and the Gouy-Chapman capacitance $C_{\mathrm{GC}} = (2\mathcal{A})(\epsilon_0\epsilon_{r,sol}/\lambda_{\mathrm{D}})\cosh\left(\frac{ze\Delta\phi_{sol}}{2k_{\mathrm{B}}T}\right)$ of the counter-charge layer in the solvent region~\cite{2012_PhysRevB_Letchworth-Weaver_Arias}, so $C^{-1} = C_g^{-1} + C_{\mathrm{GC}}^{-1}$. Even for a small gap width $t_g$, the gap capacitance $C_{g}$ can become limiting and determining for the overall interface capacitance if $C_{\mathrm{GC}}$ is sufficiently large. This is the case for small values of the Debye length $\lambda_{\mathrm{D}}$ of few $\mathrm{\AA}$ in a high-dielectric-constant solvent, as used in our calculations and consistent with typical experimental conditions, so $C \approx C_g$. For increasing Debye length, the influence of $C_{\mathrm{GC}}$ causes a decrease of $C$, as confirmed by the computational results for the Pt(111) slab in contact with implicit water solvent shown in Figure~\ref{fig_cap_vs_ts}. 

For the HBM, the counter-charge in the implicit solvent (electrolyte) region is homogeneous, so its extension is directly controlled by the width $t_{sol}$ of the solvent region, indicated in Figure~\ref{fig_scheme_models}(a). Solving Poisson's equation for the homogeneous charge distribution in the solvent region, see Supporting Information, we readily obtain a corresponding capacitance $C_{sol} = (2\mathcal{A})(\epsilon_0\epsilon_{r,sol})/(t_{sol}/4)$. We note that $t_{sol}/4$ represents the center-of-mass location of the homogeneous background charge in the implicit solvent region \emph{per half cell}. Thus, for small $t_{sol}$, the solvent-region capacitance is large, and the gap capacitance $C_g$ dominates the HBM interface capacitance, $C_{s} \approx C_g$. At larger $t_{sol}$, the influence of $C_{sol}$ causes a decrease of $C_{s}$ and, since $C_{\Omega \Phi} \approx C_{s}$, also of $C_{\Omega \Phi}$. This is confirmed by the computational results presented in Figure~\ref{fig_cap_vs_ts}. 

We thus find a direct correspondence between the Debye length $\lambda_{\mathrm{D}}$ of the PBM and the quarter of the solvent-region width $t_{sol}/4$ for the HBM, with good agreement between the corresponding capacitances. In particular, we note that for typically chosen computational parameters, i.e. $\lambda_{\mathrm{D}}  \lesssim 5\,\mathrm{\AA}$ for PBM and $t_{sol} \lesssim 20\,\mathrm{\AA}$ for HBM, both methods yield essentially equal results for the charging characteristics of the interface, which are dominated by the gap capacitance $C_g$ that is defined by the implicit solvent model, common to both methods. 

Also shown in Figure~\ref{fig_cap_vs_ts} is a fit of the HBM capacitance curve according to an electrostatic model including the presence of the gap, which is derived in the Supporting Information. The resulting Eq.~(S13) for $C_s$ was inserted into Eq.~\eqref{eq_C_Omega_Phi_HBM_model}, noting that $\eta^0_s = e^2/C_s^0$, cf. Eq.~\eqref{eq_C_surface_HBM_model}, and $\alpha_b/\alpha_s = t_b/t_s$ with $t_s = t_{sol} + 2 t_g$, to obtain $C_{\Omega \Phi}^0$ expressed in terms of $t_{sol}$ (for fixed $t_g$ and $t_b$). The gap capacitance $C_g$ and the chemical hardness $\eta_b^0$ of the bulk electrode region were fitted to the computed $C_{\Omega \Phi}^0(t_{sol})$ data\footnote{The solvent-region width was computed from $t_{sol}=\alpha_s^{sol}T$, where $T$ is the total width of the simulation cell, see Figure~\ref{fig_scheme_models}(a). The electrode slab thickness $t_b$ was computed and fixed as the distance between the Pt surface atoms (including atomic radius) of the two slab surfaces. The gap width was computed from $2 t_g = T - t_b - t_{sol}$, yielding a fixed value of $t_g = 1.17\,\mathrm{\AA}$.}, while neglecting the interaction term by setting $\eta^0_{bs} = 0$. The fitted curve reproduces the observed trend in the HBM-DFT results, demonstrating that the electrostatic model captures the essential features of the interface for the HBM. Importantly, the fitted value for the bulk chemical hardness of $\eta_b^0 = 0.047\,\mathrm{eV}$ is in good agreement with the computed hardness of $0.052\,\mathrm{eV}$ for the Pt bulk in absence of the interspace region, see above. This result corroborates the validity of the HBM model leading to Eq.~\eqref{eq_C_Omega_Phi_HBM_model}, in general, and the assumption of negligible $\eta^0_{bs}$, in particular. For the gap capacitance, a fitted value of $C_g = 0.44\,\mathrm{e\,V^{-1}}$ ($13.09\,\mathrm{\mu F\,cm^{-2}}$) was found, in good agreement with the overall PBM and HBM capacitance values in the range $\lambda_{\mathrm{D}}  \lesssim 5\,\mathrm{\AA}$ and $t_{sol} \lesssim 20\,\mathrm{\AA}$, respectively, which further supports the conclusion that the gap capacitance is the decisive factor in both HBM and PBM. As discussed above, the gap capacitance is determined by the VASPsol parameter $n_c$ that controls the distance of the implicit solvent boundary from the electrode surface atoms, and thus the width of the gap. Beyond this numerical origin and in light of its decisive role, the question for the physical meaning of the gap capacitance is emphasized~\cite{2012_PhysRevB_Letchworth-Weaver_Arias}.

\section{Conclusion}

By including the homogeneous background contribution in the definition of the grand potential, we developed an advanced homogeneous background method (HBM) that can be applied without any energy corrections and without requiring knowledge of the ``active'' fraction of excess electrons. This strongly facilitates the practical use of the HBM and extends its range of application for systems where the precise definition of the electrode--electrolyte boundary is difficult, e.g. in presence of surface adsorbates or explicit solvent molecules. We found that the advanced HBM and the Poisson-Boltzmann model (PBM) provide essentially equal results as long as the gap between the electrode surface atoms and the boundary of the implicit solvent region dominates the interface capacitance. This is fulfilled for practically chosen computational parameters, i.e. implicit-solvent-region width $t_{sol} \lesssim 20\,\mathrm{\AA}$ and $\lambda_{\mathrm{D}}  \lesssim 5\,\mathrm{\AA}$ (for PBM), in high-dielectric-constant solvents, such as water or ethylene carbonate. For the HBM, the chemical hardness of the bulk electrode region must be negligible, restricting its applicability to metal electrode materials. Under these prerequisites, the HBM provides a practical method for DFT simulations of charged electrochemical interfaces that does not require any particular implementation of a counter-ion model within the electrolyte region.

\begin{suppinfo}
Electrostatic model for the interface capacitance within the homogeneous background method (HBM).
\end{suppinfo}

\begin{acknowledgement}
T.B. acknowledges financial support in the form of a research fellowship grant funded by the SNSF (Swiss National Science Foundation). A.H and J.-S.F. thank the French National Research Agency for its support through the Labex STORE-EX Project (ANR-10LABX-76-01). This work was performed using HPC ressources from GENCI-CINES (Grant 2021-A0100910369).
\end{acknowledgement}


\providecommand{\latin}[1]{#1}
\makeatletter
\providecommand{\doi}
  {\begingroup\let\do\@makeother\dospecials
  \catcode`\{=1 \catcode`\}=2 \doi@aux}
\providecommand{\doi@aux}[1]{\endgroup\texttt{#1}}
\makeatother
\providecommand*\mcitethebibliography{\thebibliography}
\csname @ifundefined\endcsname{endmcitethebibliography}
  {\let\endmcitethebibliography\endthebibliography}{}

\end{document}